# Electrical transport between epitaxial manganites and carbon nanotubes

*Luis E. Hueso* (1) *, *Gavin Burnell* (1), *Seung Nam Cha* (2), *Jae Eun Jang* (2), *José L. Prieto* (1), *Leticia P. Granja* (1), *Christopher Bell* (1), *Dae-Joon Kang* (3,4), *Manish Chhowalla* (5), *Gehan A.J. Amaratunga* (2) *and Neil D. Mathur* (1)

(1) Department of Materials Science, University of Cambridge, Pembroke Street, Cambridge CB2 3QZ, UK

(2) Department of Engineering, University of Cambridge, Trumpington Street, Cambridge CB2 1PZ, UK

(3) Nanoscience Centre, University of Cambridge, JJ Thomson Ave, Cambridge CB3 0FF

(4) Department of Physics, Sungkyunkwan University, Suwon 440-746, Korea

(5) Ceramic and Materials Engineering, Rutgers University, Piscataway NJ 08854

* Corresponding author. E-mail: leh29@cam.ac.uk

**The possibility of performing spintronics at the molecular level may be realized in devices that combine fully spin polarized oxides such as manganites with carbon nanotubes. However, it is not clear whether electrical transport between such different material systems is viable. Here we show that the room temperature conductance of manganite-nanotube-manganite devices is only half the value recorded in similar palladium-nanotube-palladium devices. Interestingly, the former shows a pseudogap in the conductivity below the relatively high temperature of 200 K. Our results suggest the possibility of new spintronics heterostructures that exploit fully spin polarized sources and drains.**

Spin devices are one of the most sought after elements for the electronics of tomorrow. These devices aim to exploit the spin degree of freedom of the electron, in order to achieve new functionality with respect to conventional electronics. A range of effects has been observed in suitable materials and heterostructures. For example, giant magnetoresistance (GMR) is seen in metallic multilayers[1], and tunneling magnetoresistance (TMR) is seen in magnetic tunnel junctions.[2] These examples exploit the fact that ferromagnets naturally possess spin-polarized electrons. These polarized electrons are injected from one ferromagnetic region to another via non-magnetic interlayers. The relative orientation of the magnetizations of the ferromagnetic layers determines the resistance of the device. Since these relative orientations are sensitive to an external magnetic field, the devices display magnetoresistance (MR). The magnitude of the MR increases when the spin polarization ($P$) of the ferromagnetic layers is increased[3], where P is defined to be the relative difference between the populations of spin up and spin down conduction electrons. This has been investigated experimentally with materials such as $Fe_3O_4$, $CrO_2$ and, more extensively, mixed-valent manganites such as $La_{0.7}Ca_{0.3}MnO_3$.[4-7] These compounds have all been argued[8] to be half-metallic ($P$=100%) at low temperatures, such that the minority carrier band is empty. If we consider that Co is the best elemental ferromagnet with $P$=45%,[9] then we can understand the advantage of building spin devices from half-metallic materials.

In parallel with these developments in spintronics, the ever-increasing drive towards miniaturization in the semiconductor industry has delivered devices with characteristic length scales that are comparable with the dimensions of some organic molecules.[10] To go beyond the silicon roadmap,[11] there have been many recent efforts to integrate molecules into electronic circuits. In this molecular electronics approach, many molecules have been tested in order to establish whether they are able to perform basic electronic processes (such as amplification or rectification), or even new and unexpected effects.[12] Out of all possible molecular building blocks, carbon nanotubes (CNTs) are among the most studied and have shown the most promising results.[13] For example, logic gates[14] have been fabricated using room-temperature transistors based on semiconducting single-walled CNTs.[15] Also, ballistic conduction in nanotubes, which leads to extraordinary properties, offers the possibility of faster devices that consume less power.[16]

There have been some attempts at integrating spintronics and molecular electronics: Tsukagoshi *et al.* successfully contacted multiwalled CNTs to ferromagnetic Co contacts, and showed a few percent low-temperature MR effect.[17] Other groups have obtained similar but inconsistent results.[18-21]

Our purpose is to show the possibility of electrically connecting a molecule to a magnetic material that shows full spin polarization. We have chosen to work with carbon nanotubes and a mixed valent manganese oxide (a manganite).

If electrical conductivity is possible between such different materials systems, then it may be possible to control electronic spin states in carbon nanotubes over distances that are long compared with tunnel junction barrier widths of a few nanometers. Moreover, the diameter of a carbon nanotube is much smaller than the lateral dimensions of a typical barrier, suggesting improved performance. Thus one can envisage large MR effects, as well as the creation of extremely localized sources of spin-polarized electrons for more challenging applications such as quantum computing.[22] From a more fundamental point of view, ferromagnetic contacts may prove a useful probe of spin-charge separation in Luttinger-liquid (LL) materials.[23-25]

High quality single crystal films of $La_{2/3}Sr_{1/3}MnO_3$ (LSMO) were grown on $SrTiO_3$ (001) substrates by pulsed laser deposition. This particular manganite composition was chosen because it is fully spin polarized at low temperature [8] and its ferromagnetic properties persist well above room temperature.[26] Commercially available $SrTiO_3$ (001) substrates (Crystal GmbH, Germany) were chemically etched with a mixture of $HF/NH_4F$ prior to the deposition in order to obtain an atomically flat surface.[27] A KrF laser (repetition rate 1 Hz, fluence 2.0 J/cm$^2$) was used to ablate a polycrystalline LSMO target (Praxair, USA) such that the plume impinged substrates positioned 7.5 cm below. The films were grown at a substrate temperature of 750 $^0$C in a flowing oxygen ambient of 15 Pa. After deposition, the films were annealed at the same temperature in 60 kPa oxygen for 1 hour. X-ray diffraction (XRD) showed that the films are coherently strained and are 30 nm thick. This thickness represents a compromise between the need to maintain the intrinsic properties of the manganite film, and the need to define manganite electrodes without milling deep trenches into which the connecting nanotubes would hang. Using atomic force microscopy (AFM), we observed unit cell high terraces (Fig. 1) confirming that step flow growth took place and therefore that the film is epitaxial.

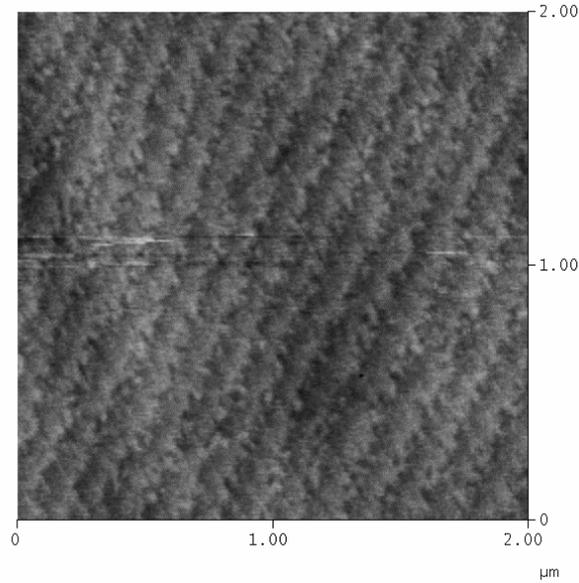

**Figure 1.** AFM image of an LSMO film used in this study. The 3.7 Å terraces are evidence of good epitaxy, given that the underlying substrate possesses similar terraces due to the vicinal off-cut.

Pd was chosen as a control electrode material because it combines a high work-function with good wetting properties for nanotubes, leading to a low contact resistance.[28] Pd films were grown on $SrTiO_3$ (001) substrates by d.c. sputtering in a dedicated chamber for non-magnetic materials to avoid ferromagnetic contamination of the Pd. The Pd resistivity was found to be 20 μΩ.cm at 10 K.

Lines of width 1 μm and separation 1.5 μm were created along [100] in the LSMO and Pd films by conventional photolithography and $Ar^+$ ion milling. An over-milling of 5 nm ensured no electrical contact between the conducting lines. Commercial high-quality multiwalled CNTs (Iljin Nanotech Co. Ltd., Korea) with lengths in excess of 3 μm were dispersed on the patterned chips from an ultrasonicated solution. Nanotubes making electrical contact between LSMO electrodes were imaged by scanning electron microscopy (Fig. 2). No further processing, e.g. annealing or e-beam exposure, was performed in order to improve the contact resistance. We found that *in situ* CNT growth was not viable because the reducing high temperature environment degrades the ferromagnetism of the LSMO catastrophically.

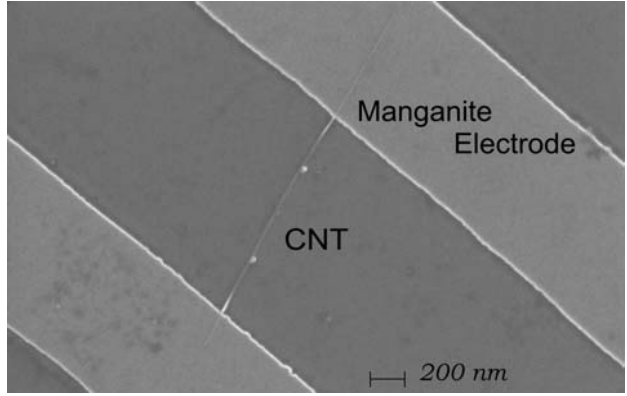

**Figure 2.** SEM photograph of a multiwalled carbon nanotube in electrical contact with two manganite lines that act as electrodes for the LSMO-CNT-LSMO devices.

We now discuss the electrical transport and magnetic properties of the LSMO films prior to patterning. Metallic behavior was seen from room temperature down to the lowest temperature measured, and a residual resistivity of around 50 μΩ.cm was recorded at 10 K (Fig. 3). Ferromagnetic behavior was seen below a Curie temperature of 360 K, and the low temperature saturation magnetization is 3.6 $\mu_B$/Mn (Fig. 3 (inset)), just 3% less than the theoretical value of 3.7 $\mu_B$/Mn. Therefore the manganite films are of the highest possible quality[29] ensuring a high *P*.

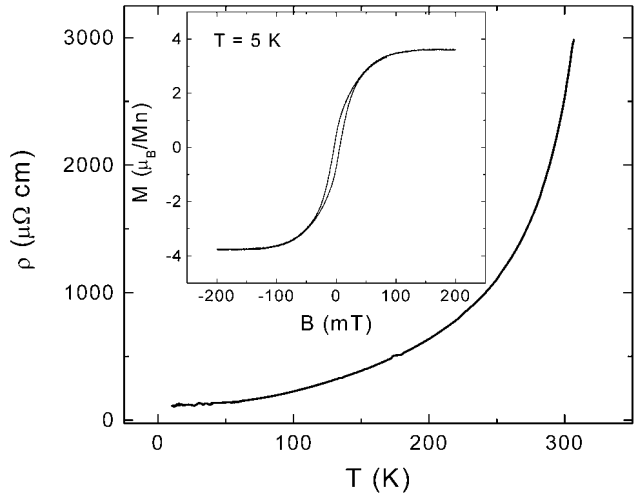

**Figure 3.** Electrical and magnetic data for the [100] direction of an LSMO film. Resistivity *ρ* versus temperature *T* and (inset) magnetization *M* versus applied magnetic field *B*. The metallic behavior of the resistivity below room temperature and the 97% saturation of the low-temperature magnetization demonstrate optimal quality.

We now present results for the Pd-CNT-Pd control devices. A bias voltage (V) was applied between the source and drain Pd electrodes, and the current (I) was measured. Non-linearity in the I-V characteristics persisted up to room temperature (Fig. 4), and the zero-bias anomaly (ZBA) is clearly enhanced at lower temperatures (Fig. 4 (inset)). Although we find no sign of blocked conductivity at low bias at the lowest temperature studied (4.2 K), we attribute this to the fact that we only took data down to 4.2 K. The conductance was found to vary with the characteristic power laws $G(T) \propto T^\alpha$ ($eV \ll kT$) and $G(V) \propto V^\alpha$ ($eV \gg kT$) with $\alpha \sim 0.4$ — not shown. Similar behavior has also been observed in multiwalled CNT-metal samples, and explained using the environmental Coulomb Blockade theory.[30] Note that we rule out the possibility of conduction pathways via the substrate, since in similar measurements of Pd lines without CNTs, the current is four orders of magnitude smaller (<1 nA at 1 V).

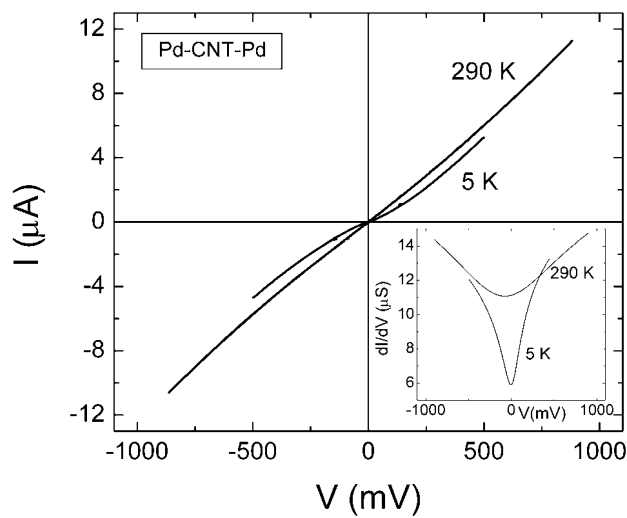

**Figure 4.** A typical current (I)-voltage (V) characteristic for Pd-CNT-Pd devices. **Inset**: A typical dI/dV trace showing the development of a zero-bias anomaly at low bias.

We now turn to our LSMO-CNT-LSMO devices. Our first observation is that all devices measured behave similarly, as do the Pd-CNT-Pd devices. At room temperature, the conductance of the Pd-CNT-Pd devices is around double the conductance of the LSMO-CNT-LSMO devices. Given the resistivities of the device components, the CNT contact resistance in both types of device (which we cannot measure in our two-point geometry) must dominate. Therefore the LSMO-CNT contact resistance is roughly double the Pd-CNT contact resistance.

The conductance of the LSMO-CNT-LSMO devices decreases with decreasing temperature, and becomes very small ($<5\times10^{-7}$ S) at low bias at ~200 K (Fig. 5). This is a remarkably high temperature compared with typical values (~10 K)[13] in nanotube devices. The pseudogap saturates to ~250 meV at very low temperatures (Fig. 6). This energy is remarkably large with respect to typical values (5-25 meV)[13] in nanotube devices. At 5 K, the conductance in the high voltage limit again follows a $G(V) \propto V^\alpha$ power law, but with $\alpha \sim 2.8$ in all samples studied (Fig. 6 inset). This value is far from the value of $\alpha \sim 0.4$ obtained earlier for Pd-CNT-Pd devices, and would imply strong repulsive interactions in LL theory.

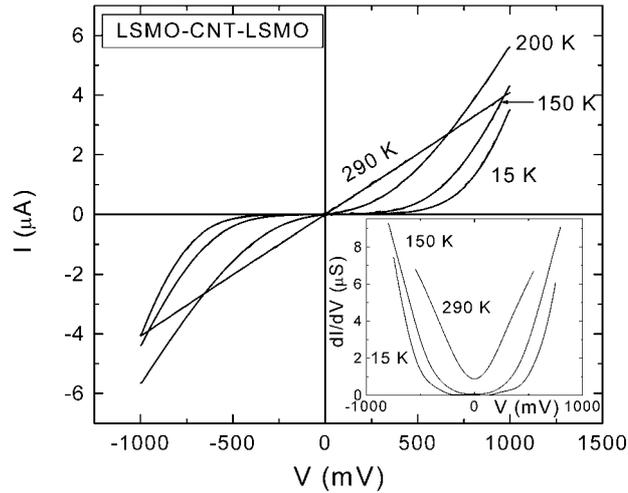

**Figure 5.** Typical I-V characteristics at different temperatures for the LSMO-CNT-LSMO devices. As the temperature is lowered, a region of very small conductance appears al low voltages. This can be clearly seen in the dI/dV curves (inset).

The presence of the pseudogap in all of our LSMO-CNT-LSMO devices suggests that interfacial details are independent of the details of at least the outermost nanotube (e.g. diameter, chirality, etc.). For example in every case, the crystal lattice mismatch could be poor, and the LSMO surface magnetization[30] and thus conductivity suppressed. We note that our device performance is reminiscent of two back-to-back

Schottky barriers, but the complexities discussed above preclude the standard analysis.

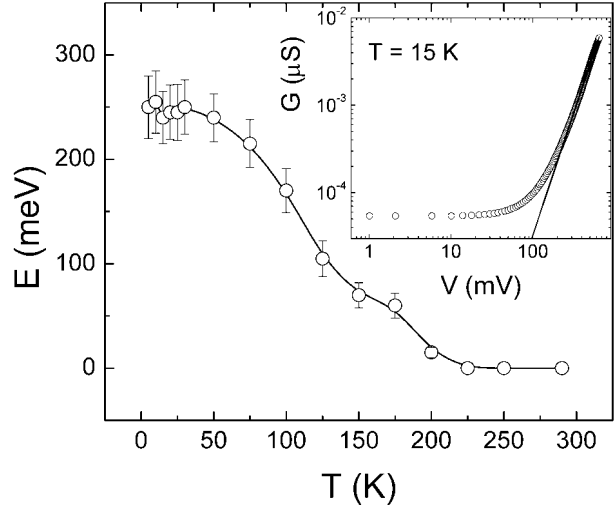

**Figure 6.** Estimated values of the pseudogap energy (E) extracted from dI/dV for the LSMO-CNT-LSMO devices. The line is a guide to the eye. **Inset:** A typical low temperature conductance $G$ as a function of the applied voltage $V$ for the LSMO-CNT-LSMO devices. The line represents a power-law fit with $\alpha=2.8$.

In conclusion, we have established that it is possible to pass an electrical current between a half-metallic oxide and a carbon nanotube. At room temperature, the contact resistance is only double the value recorded for similar devices with non-magnetic metallic electrodes made from Pd. Below 200 K, a gap develops in the I-V characteristic of the manganite-nanotube devices. In the Pd-nanotube devices we have seen that the zero-bias conductance decreases down to our base temperature of 4.2 K, and we anticipate the development of a gap at lower temperatures. Future device improvements include the incorporation of a gate electrode to permit the carrier density in semiconducting tubes to be varied, the use of LSMO lines with different widths to achieve different coercive fields, and the use of ultra thin "wetting" layers for reduced contact resistance.

We thank Peter Littlewood, Maria Calderón, Valeria Ferrari, Miguel Pruneda and Emilio Artacho for helpful discussions. This work was funded by the UK EPSRC, the EU Marie Curie Fellowship (LEH), the Royal Society and the NSF (MC).

**References**

(1) Baibich, M.N.; Broto, J.M.; Fert, A.; Nguyen Van Dau, F.; Petroff, F.; Eitenne, P.; Creuzet, G.; Friederich, A.; Chazelas, J. *Phys. Rev. Lett.* **1988**, *61*, 2472.

(2) Moodera, J.S.; Kinder, L.R.; Wong, T.M.; Meservey, R. *Phys. Rev. Lett.* **1995**, *74*, 3273.

(3) Julliere, M. *Phys. Lett. A* **1975**, *54*, 225.

(4) Li, X.W.; Gupta, A.; Xiao, G.; Qian, W.; Dravid, V.P. *Appl. Phys. Lett.* **1998**, *73*, 3282.

(5) Gupta, A.; Li, X.W.; Xiao, G. *Appl. Phys. Lett.* **2001**, *78*, 1894.

(6) Jo, M.-H.; Mathur, N.D.; Todd, N.K.; Blamire, M.G. *Phys. Rev. B* **2000**, *61*, R14905.

(7) Bowen, M.; Bibes, M.; Barthelemy, A.; Contour, J.-P.; Anane, A.; Lemaitre, Y.; Fert, A. *Appl. Phys. Lett.* **2003**, *82*, 233.

(8) Park, J.-H.; Vescovo, E.; Kim, H.-J.; Kwon, C.; Ramesh, R.; Venkatesan, T. *Nature* **1998**, *392*, 794.

(9) Meservey, R.; Tedrow, P.M. *Phys. Rep.* **1994**, *238*, 175.

(10) International Technology Roadmap for Semiconductors, 2004 Update. (Available at http://public.itrs.net/)

(11) Mathur, N.D. *Nature* **2002** *419*, 573.

(12) Heath, J.R.; Ratner M.A. *Physics Today* **2003**, *56*, 43.

(13) *Carbon Nanotubes: Synthesis, Structure, Properties and Applications*; Dresselhaus, M., Dresselhaus, G., Avouris, Ph., Eds.; Springer-Verlag: Berlin, 2001.

(14) Bachtold, A.; Hadley, P.; Nakanishi, T.; Dekker, C. *Science* **2001**, *294*, 1317.

(15) Tans, S.J.; Verschueren, A.R.M.; Dekker, C. *Nature* **1998**, *393*, 49.

(16) Kong, J.; Yenilmez, E.; Tombler, T.W.; Kim, W.; Dai, H.; Laughlin, R.B.; Liu, L.; Jayanthi, C.S.; Wu, S.Y. *Phys. Rev. Lett.* **2001**, *87*, 106801.

(17) Tsukagoshi, K.; Alphenaar, B.W.; Ago,H. *Nature* **1999**, *401*, 572.

(18) Orgassa, D.; Mankey, G.J.; Fujiwara, H. *Nanotechnology* **2001**, *12*, 281.

(19) Zhao, B.; Monch, I., Muhl, T.; Vinzenbelrg, H.; Schneider, C.M. *J. Appl. Phys.* **2002**, *91*, 7026.

(20) Kim, J.-R.; So, H.M.; Kim, J.-J.; Kim, J. *Phys. Rev. B* **2002**, *66*, 233401.

(21) Jensen, A.; Hauptmann, J.R.; Nygård, J.; Sadouski, J.; Lindelof, P.E.; *Nano Lett.* **2004**, *4*, 349.

(22) Hueso, L.; Mathur, N. *Nature* **2004**, *427*, 301.

(23) Si, Q. *Phys. Rev. Lett.* **1998**, *81*, 3191.

(24) Mehrez, H.; Taylor, J.; Guo, H.; Wang, J.; Roland, C. *Phys. Rev. Lett.* **2000**, *84*, 2682.

(25) Balents, L.; Egger, R. *Phys. Rev. Lett.* **2000**, *85*, 3464.

(26) Urushibara, A.; Moritomo, Y.; Arima, A.; Asamitsu, A.; Kido, G.; Tokura, Y. *Phys. Rev. B* **1995**, *51*, 14103.

(27) Koster, G.; Kropman, B.L.; Rijnders, G.J.H.M.; Blank, D.H.A.; Rogalla, H. *Appl. Phys. Lett.* **1998**, *73*, 2920.

(28) Javey, A.; Guo, J.; Wang, Q.; Lundstrom, M.; Dai, H.; *Nature* **2003**, *424*, 654.

(29) Pellier, W.; Lecoeur, Ph.; Mercey, B. *J. Phys.: Condens. Matter* **2001**, *13*, R915.

(30) Bachtold, A.; de Jonge, M.; Grove-Rasmussen, K.; McEuen, P.L.; Buitelaar, M.; Schonenberger, C. *Phys. Rev. Lett.* **2001**, *87*, 166801.

(31) Park, J.-H.; Vescovo, E.; Kim, H.J.; Kwon, C.; Ramesh, R.; Venkatesan, T. *Phys. Rev. Lett* **1998**, *81*, 1953.